\documentclass[
 aip,amssymb,jcp,
reprint,floatfix,
]{revtex4-2}
\usepackage{graphicx}
\usepackage{color}
\usepackage{hyperref}
\usepackage{float}
\usepackage{array, multirow}
\usepackage{amsmath}
\usepackage{newtxtext}
\usepackage{booktabs}
\usepackage{xcolor}
\usepackage{soul}
\usepackage{url}
\usepackage{gensymb}
\usepackage[T1]{fontenc}
\usepackage[caption=false]{subfig}

\hypersetup{colorlinks,
    linkcolor={blue!75!black!80!yellow},
    citecolor={blue!75!black!80!yellow},
    urlcolor={blue!75!black!80!yellow}
    }

\begin{document}

\title{Understanding the effects of competing spin-pair dephasing pathways in molecular spins}

\author{Timothy J. Krogmeier}
\affiliation{Department of Chemistry, University of Minnesota, Minneapolis, MN 55455 USA}
\author{James Bradley}
\affiliation{Department of Chemistry, University of Minnesota, Minneapolis, MN 55455 USA}
\author{Anthony W. Schlimgen}
\affiliation{Department of Chemistry, University of Minnesota, Minneapolis, MN 55455 USA}
\author{Kade Head-Marsden}
\affiliation{Department of Chemistry, University of Minnesota, Minneapolis, MN 55455 USA}
\email{khm@umn.edu}

\begin{abstract}
Molecular spins offer promise in emerging quantum technologies such as quantum sensing and computing. At low temperatures, nuclear spin-spin interactions affect electron spin coherence lifetimes through pure dephasing. Nuclear-spin noise can originate from spin pairs within a molecule itself, pairs in a surrounding environment system, or pairs in which one spin is on the molecule and the other in the environment. Improving coherence times requires detailed knowledge of the dominant sources of dephasing. Here, we analyze the decoherence behavior of two molecular qubit candidates with various ligands and in different nuclear-spin containing solvents. We apply an electronic-structure enhanced, non-Markovian perturbative theoretical method to connect experimentally comparable dephasing times to individual spin pairs. This analysis allows the development of a computational workflow to strategically improve coherence lifetimes in spin systems where decoherence is dominated by spin-spin dephasing. 
\end{abstract}

\maketitle

\section{Introduction}

Molecular spins have been proposed for use in quantum information science (QIS) as qubits and quantum sensors.~\cite{Gaita:2019,Bayliss:2020,Wasielewski:2020a,Yu:2021, Mullin:2023a,Moreno:2025} The efficacy of molecular spins in quantum technologies depends on many properties, in particular on maintaining superposition lifetimes long enough for gate operations and state-preparation protocols. At low temperatures, the major contributor to the decoherence time ($T_2$) of a superposition state is the dephasing time ($T_\phi$) due to nuclear-spin noise.~\cite{Ramsey:1953a,Lazzeretti:2003a,Purcell:1952a,Sham:2006a,Zhao:2012} Significant research has focused on improving coherence times using different molecular motifs,~\cite{Ghosh:2026,Martinez:2025,Boning:2026,Baldinelli:2025} including experimental work synthesizing molecular qubit candidates without spin-active nuclei in the ligands and using deuterated solvents.~\cite{Yu:2016,Zadrozny:2015,Mkami:2014,Qiu:2023,Bader:2017} However, eliminating spin-active nuclei from synthetic motifs restricts the design space of the ligands, and many potential QIS technologies occur in the presence of solvents that may contain spin-active nuclei even after partial deuteration.~\cite{Ghirri:2017,Coronado:2020} Taken together, the molecular ligand structure and the solvent environment provide pathways for significant decoherence of molecular spins, which is unlikely to be eliminated through partial exclusion of spin-active nuclei. Alternatively, molecular modifications have been explored as an additional route for modulating spin-dephasing times.~\cite{Graham:2017,Fataftah:2019,Jackson:2019,Dai:2018,Bayliss:2022} Predicting which solvent or structure changes are most important is challenging due to the interplay of electronic structure and quantum dynamics in these experiments. This challenge has inspired the development of computational techniques that can delineate contributions to spin dephasing to aid in the improved design of quantum spin technologies.

Early experimental studies showed improvements to the electron coherence time for a variety of transition metal systems.~\cite{Graham:2017,Fataftah:2019,Shiddiq:2016,Schafter:2023,Maylander:2023,Atzori:2016,Atzori:2018,Ariciu:2019} A more recent study has shown that modifications of a molecular spin system with nuclear spin-bearing ligands do not appreciably alter the dephasing time of the electronic spin.~\cite{Akintola:2026} This leaves an open question as to when molecular modifications can impact coherence lifetimes versus when the lifetime is strictly solvent limited. While experiments have made significant progress in this area, this understanding could be facilitated by careful theoretical and computational analysis of the microscopic contributions to spin dephasing, considering both the molecular electronic structure and the time-dependent spin-dynamics of molecular spins in realistic environments. Previous theoretical results have considered the effects of spin-spin interactions on the electron coherence, for example, increasing solvent deuteration was shown to lengthen the coherence time among various molecular motifs.~\cite{Jahn:2024,Jahn:2022,Canarie:2020,Chen:2020,Onizhuk:2024,Onizhuk:2025,Ryan:2025}
\begin{figure}[h!]
    \centering
    \includegraphics[width=\linewidth, trim = 0cm 18cm 8cm 0cm]{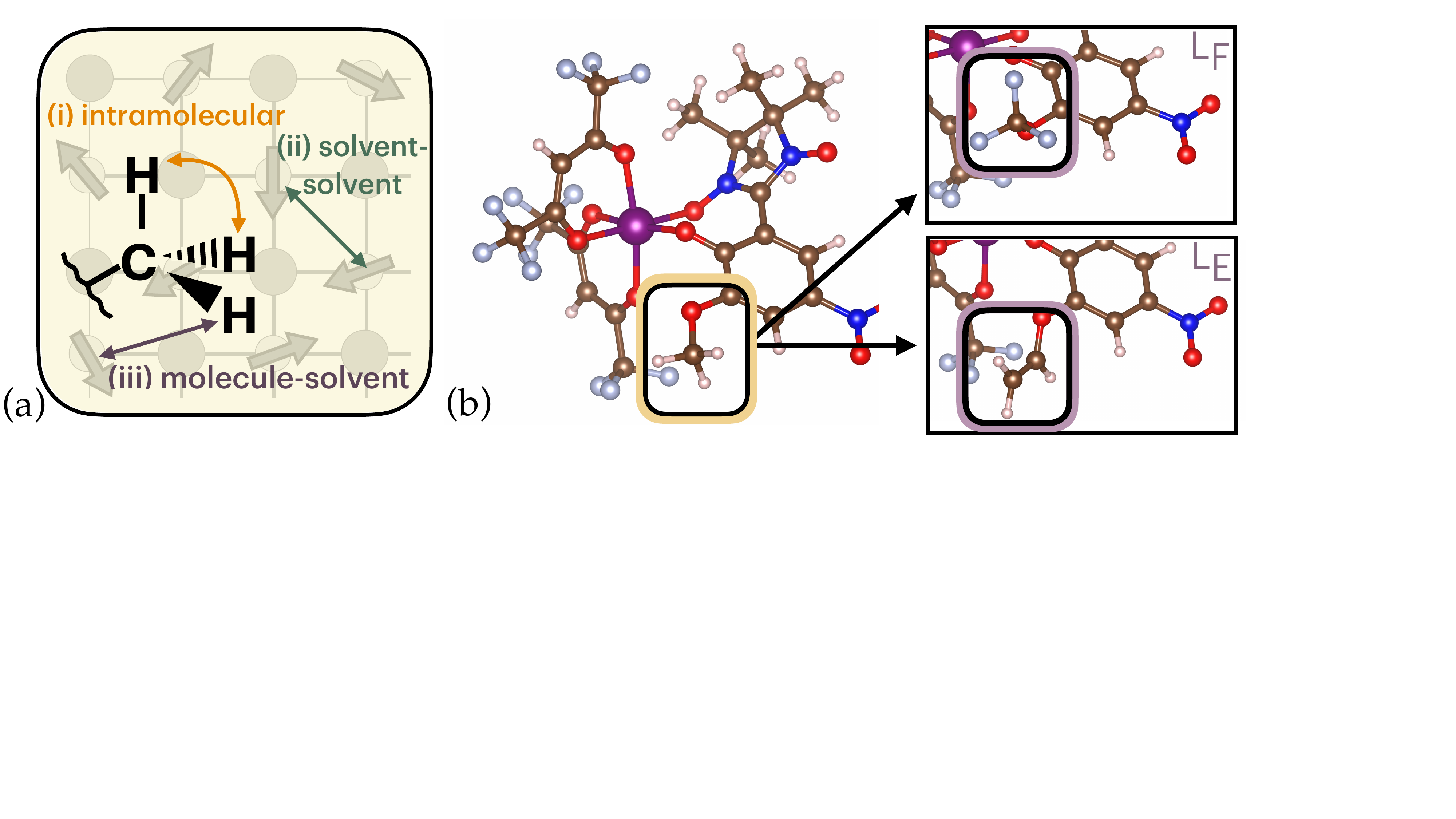}
    \caption{(a) A methyl group as part of a ligand on a transition-metal centered molecule, where the spin-spin interactions are highlighted for the intramolecular (yellow), solvent-solvent (green), and molecule-solvent (purple) interactions. (b) [\textcolor{black}{Zn}L]$^-$ structure where the methoxy methyl group, which was determined to contain spin pairs with the highest contribution to intramolecular spin dephasing, is highlighted, along with two proposed modifications replacing the methyl hydrogens with fluorines (L$_F$) and replacing the methyl group with ethylene (L$_E$). Fluorine is shown in light blue, nitrogen in blue, hydrogen in white, carbon in brown, oxygen in red, and \textcolor{black}{zinc in purple}.}
    \label{fig:schematic}
\end{figure}
 Identifying improvements in molecular spins requires the treatment of spin-spin interactions between pairs of spin-active nuclei in the molecule and solvent environment. In general, we can distinguish the interaction of nuclear spins with an electron spin by contributions from (i) intramolecular-spin pairs, (ii) solvent-solvent nuclear-spin pairs, and (iii) molecule-solvent pairs, shown by yellow, green, and purple arrows, respectively, in Figure~\ref{fig:schematic}~(a). The nuclear-spin contributions additionally depend on the distance of the nuclear-spin pair from the electron spin, due to the strong hyperfine coupling to the electron, which can be predicted with electronic structure theory. Both theory and experiment have demonstrated the presence of the \emph{spin-diffusion barrier}, in which nuclear-spin pairs close to an electron, approximately 4~\AA ~away, are prevented from exchanging magnetization.~\cite{Graham:2017,Chen:2020,Krogmeier:2024,Krogmeier:2026,Guichard:2015} Identifying and proposing molecular modifications to improve coherence times therefore requires careful treatment of the underlying electronic structure, distinguishing between the contributions of different spin pairs to dephasing, and a consistent and microscopically derived dynamical description of the interactions.

Here, we demonstrate the utility of computational analysis of different electronic and nuclear spin effects on dynamical dephasing in two molecular qubit candidates, describing the influence of electronic structure, ligand substitutions, and solvent modifications. Recent experiments measured the Hahn-echo decay times at low temperatures for two molecular qubit candidates, [ZnL]$^-$ and [NiL]$^-$ where L $=$ (hfac)$_2$2-(2-hydroxy-3-methoxy-5-nitrophenyl)-4,4,5,5-tetramethyl-4,5-dihydro-1\textit{H}-imidazol-3-oxide-1-oxyl, and found that [ZnL]$^-$ had a faster dephasing time than [NiL]$^-$.~\cite{Martins:2026,Spinu:2021} The electronic structure of the two molecules is qualitatively different, with the spin density primarily on the imidazoline ligand in [ZnL]$^-$, but on the nickel ion in [NiL]$^-$. The experiments were performed in an organic solvent composed of hydrogen-containing molecules, making the dephasing times sensitive to nuclear spin flip-flops among intramolecular, solvent-solvent, and molecule-solvent spins. 

We simulate the spin-dephasing dynamics using a recently derived second-order time-convolutionless (TCL2) master equation that employs a pair approximation method,~\cite{Krogmeier:2026} yielding experimentally-comparable dynamics. This approach enables the use of density functional theory (DFT) to compute the coupling between electron and nuclear spins on the molecule and the analysis of which pairs contribute most to the dephasing. We can identify a major source of dephasing in [ZnL]$^-$ and [NiL]$^-$, denoted  concisely as ZnL and NiL respectively, and propose two ligand modifications to improve coherence times. Analysis of hyperfine interactions, and consideration of the spin-diffusion barrier motivates replacing the methoxy CH$_3$ group on the ligand (L) with a CF$_3$ or C$_2$H$_3$ (ethylene) group, shown schematically in Fig.~\ref{fig:schematic}~(b), which are denoted as L$_F$ and L$_E$, respectively. Our results explain the experimentally observed trends in these particular species, and also provide a general and practical workflow for understanding and tuning low-temperature electronic spin dephasing in spin-active environments.

\section{Theory and Methods}

\subsection{Dephasing dynamics}
The coherence of an electron spin is determined by the off-diagonal element of the reduced density matrix at time $t$ and can be described by the following expression derived through the TCL2 master equation,~\cite{Krogmeier:2026}
\begin{equation}\label{eq:TCL2-coherence}
\rho_e^{01}(t) = \rho_e^{01}(0)e^{-\sum_{kl}W_{kl}(t)},
\end{equation}
where,
\begin{equation}\label{eq:TCL2-W}
W_{kl}(t)=\left( \frac{2\Delta_{kl}b_{kl}}{\Delta_{kl}^2 + b_{kl}^2}\right)^2\sin^4{\left(\frac{t}{4}\sqrt{\Delta_{kl}^2 + b_{kl}^2}\right)},
\end{equation}
$\Delta_{kl} = A_k-A_l$ is the difference between the $zz$ components of the hyperfine tensors for nuclear spins $k$ and $l$, and $b_{kl}$ are the nuclear spin-spin dipolar coupling constants between nuclear spins $k$ and $l$. \textcolor{black}{The time dependence arises in the $\sin^4(\cdot)$ function, and $\Delta_{kl}$ and $b_{kl}$ can generally be obtained from electronic structure inputs.} There are two important features contained in the equation above: the modulation depth,
\begin{equation}\label{eq:TCL2-modulation}
    \alpha^2_{kl} = \left( \frac{2\Delta_{kl}b_{kl}}{\Delta_{kl}^2 + b_{kl}^2}\right)^2
\end{equation}
and the frequency, 
\begin{equation}\label{eq:TCL2-frequency}
    f_{kl}=\frac{1}{4}\sqrt{\Delta_{kl}^2 + b_{kl}^2}.
\end{equation}
The modulation depth is an amplitude that corresponds to the contribution of each spin pair to the echo decay, while the frequency determines the oscillation speed of the electron coherence.~\cite{Jeschke:2023} These are both key quantities in our analysis as a larger modulation depth reached within a timeframe given approximately by $\frac{1}{2f_{kl}}$ indicates increased contribution of a nuclear spin pair to electron dephasing. We note that a 4$^\textrm{th}$-order TCL equation was also derived to treat spin-spin induced electron dephasing;~
\cite{Krogmeier:2026} however, for these systems we find that the TCL2 is sufficient for the analysis of the dephasing contributions from specific intramolecular, solvent-solvent, and molecule-solvent nuclear spin pairs.

\subsection{Electronic structure}

Nuclear spin noise dephases an electron spin via the hyperfine coupling parameters $A_{zz}$ in Eq.~\ref{eq:TCL2-W}, which depends on the electron-spin density and can be determined from electronic structure calculations. We compute hyperfine tensors with DFT in six compounds: NiL, NiL$_F$,  NiL$_E$, ZnL, ZnL$_F$, and ZnL$_E$. For each of these compounds we use the B3LYP hybrid functional with density fitting implemented in the Orca 6 quantum chemistry software.~\cite{Neese:2025,Neese:2009,Neese:2003a,Weigend:2006,Becke:1988a,Parr:1988a,Chmela:2018} \textcolor{black}{Other functionals were tested for the unmodified structures and yielded comparable results, with details in the SI.} We optimize the geometries using a def2-TZVP basis set for all atoms,~\cite{Weigend:2005} then compute hyperfine tensors, including contributions from dipole-dipole interactions, spin-orbit coupling, and the Fermi contact, using the exact two-component (X2C) scalar relativistic Hamiltonian and X2C-TZVP basis set for all atoms except hydrogen and fluorine.~\cite{Franzke:2019,Pollak:2017} \textcolor{black}{For the hydrogen and fluorine atoms on the molecule we use Barone's DZ basis set, which is designed for computing hyperfine coupling constants at reasonable cost.}~\cite{Barone:2008a,Barone:2008b,Barone:2008c,Barone:2010}

\subsection{Solvent spin pairs}

We consider the surrounding solvent system to compute additional contributions to the echo decay from Eq.~\ref{eq:TCL2-W} The Hahn-echo measurements on NiL and ZnL are taken in a mixture of dichloromethane and toluene, which are solvents with spin-active nuclei.~\cite{Martins:2026} To simulate the effects of the solvent hydrogens, we place each molecule at the center of a cubic box and populate the box with randomly placed hydrogen spins, removing any solvent hydrogens that are within one \AA ~of any of the molecular atoms. We compute a density of hydrogen spins by considering the density of solvent molecules, with details shown in S1 in the Supporting Information (SI), and average the dynamics over 1000 randomly placed configurations. This approach does not account for solvent packing effects which depend on the molecule and solvent, which could be included by semi-empirical or molecular dynamics calculations; however, we find \textcolor{black}{that} the dynamics converge with our sampling approach, and are suitable for the purposes of this study. We compute interactions between the electron and solvent hydrogen spins with the point-dipole approximation, and add these contributions to the TCL2 approach. \textcolor{black}{While this involves several approximations, we find that these approximations are sufficient to capture qualitative trends for the molecules considered. Future work will be dedicated to systematically testing and improving upon these approximations for application of TCL methods when molecule-solvent and solvent-solvent spin pair interactions are significant for the dephasing dynamics.}

\section{Results}

\begin{figure}[h!]
    \centering
   \includegraphics[width=\linewidth, trim = 0cm 10.5cm 0cm 0cm, clip]{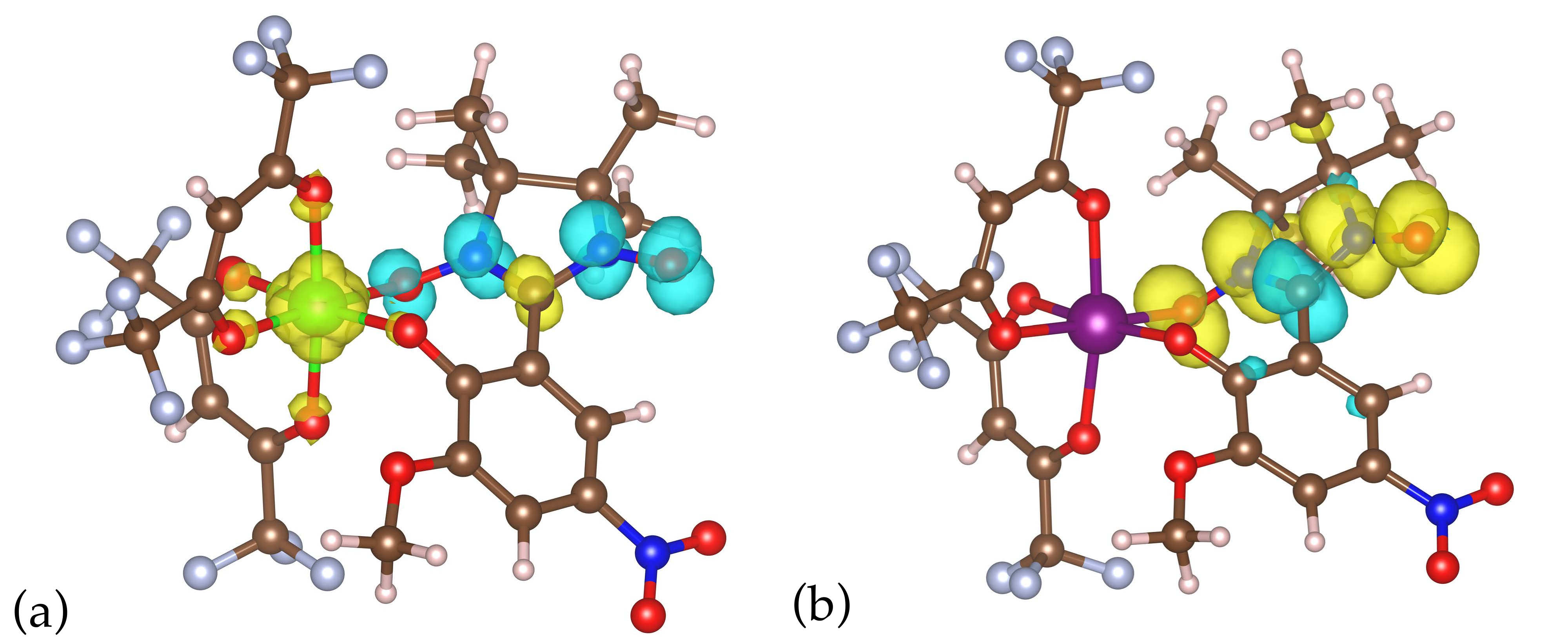}
    \caption{Electron spin density for the (a) Ni structure and (b) Zn structure. Fluorine is shown in light blue, nitrogen in blue, hydrogen in white, carbon in brown, oxygen in red, nickel in green, and zinc in purple.}
    \label{fig:spin-densities}
\end{figure}

The electronic structures of the NiL and ZnL species exhibit important qualitative differences in the spin densities, shown in Figure~\ref{fig:spin-densities}~(a) and (b), respectively. 
In the Ni case, the electron is located on the metal center, in contrast to the Zn species where the electron is located on the ligand. \textcolor{black}{While Mulliken populations are highly basis dependent, in these calculations the Ni has a spin density of 1.716 while the Zn has a mere 0.003.} We distinguish between two types of methyl groups in these structures: the methyl group in the methoxy, highlighted on the left in Fig.~\ref{fig:schematic}~(b), and the four methyl groups of the imidazoline ring. For the NiL (ZnL) structure, the methoxy hydrogens are about 5.6 \AA ~(7.8 \AA), while the four methyl groups of the imidazoline ring have hydrogens from about 5.6 \AA~ (4.1 \AA), where the distances are measured from the location of the electronic spin for each species. The difference in the electron spin density, combined with the location of the methoxy hydrogens, results in different couplings and dynamics between the two structures due to the intramolecular nuclear spin-spin interactions. The electron spin densities for all modified structures can be found in S2 in the SI. 
\begin{figure}[h!]
    \centering
    \includegraphics[width=0.8\linewidth]{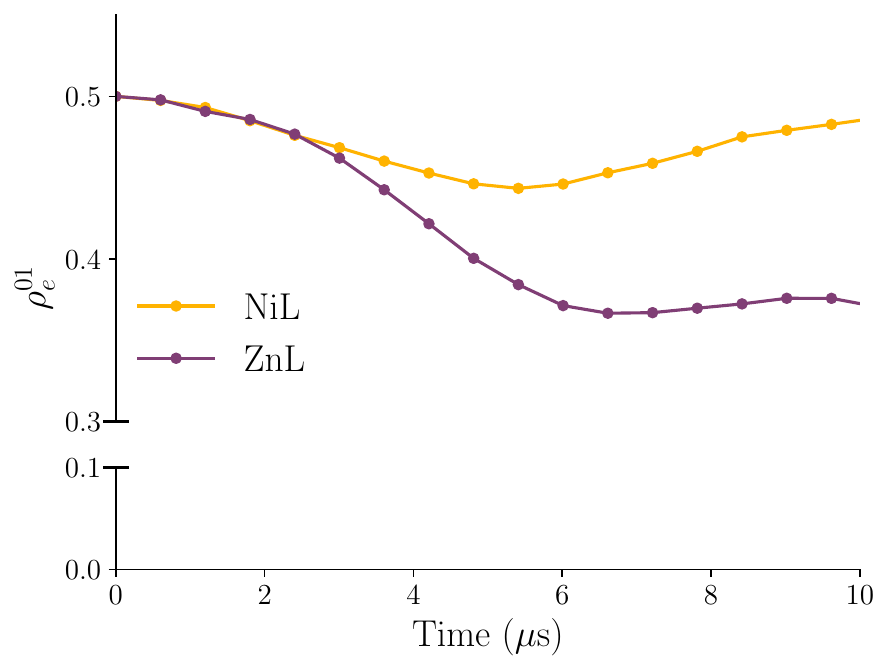}
    \caption{Electron coherence in the isolated molecules NiL and ZnL.}
    \label{fig:niZnSingleMolecule}
\end{figure}

\begin{figure}[h!]
    \centering
    \includegraphics[width=0.8\linewidth]{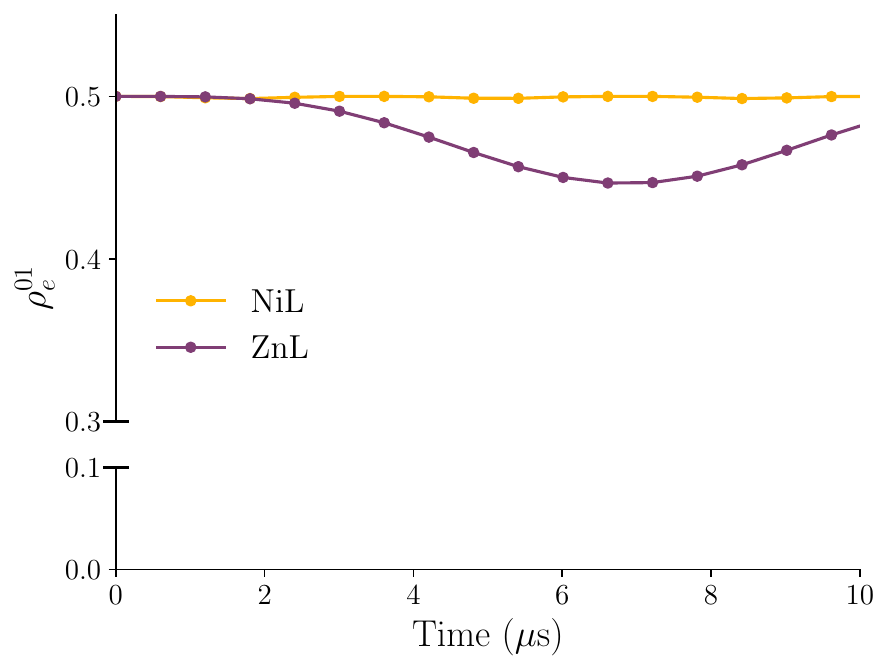}
    \caption{Electron spin coherence due to a single pair of molecular hydrogen spins, computed using Eq.~\ref{eq:TCL2-coherence}.}
    \label{fig:niZnMajorContributors}
\end{figure}

\begin{figure}[h!]
    \centering
    \includegraphics[width=0.8\linewidth, trim = 0cm 0cm 40cm 0cm, clip]{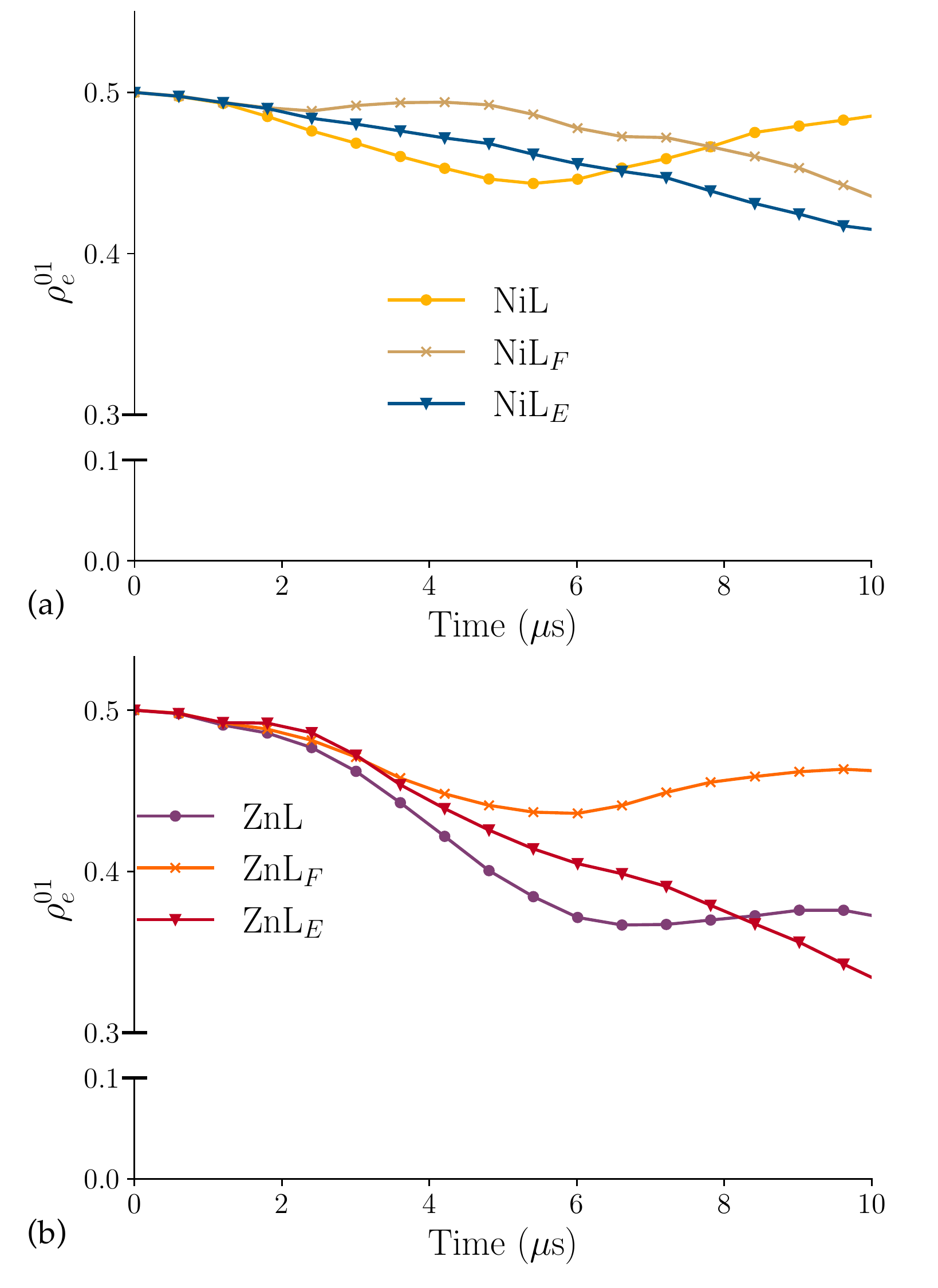}
    \caption{Electron coherence over time for (a) the unmodified NiL structure compared to the two modified structures, NiL$_F$ and NiL$_E$ and (b) the unmodified ZnL structure compared to the two modified structures, ZnL$_F$ and ZnL$_E$.}
    \label{fig:niModifiedSingleMol}
\end{figure}

\textcolor{black}{For all dynamics calculations, we start with an initial density matrix with an off-diagonal element, or coherence, of 0.5, $\rho_e^{01}(0)=0.5$. While the TCL4 framework is also accessible,~\cite{Krogmeier:2026} we compared the dynamics from TCL2 and TCL4 in the SI to confirm that TCL2 is sufficient for these molecular species and timescales.} In Figure~\ref{fig:niZnSingleMolecule} we show the spin dephasing dynamics of NiL and ZnL when including only the intramolecular spin interactions, i.e. in a solvent-free box. Under these conditions, NiL has a slower onset of decay in coherence than ZnL, even though the molecular structures are only marginally different. The difference in these dynamics results from the different electron spin densities in NiL and ZnL, and their interaction with the methoxy hydrogens. Specifically, the nickel-centered electron in NiL is closer to the methoxy hydrogens than the electron in ZnL, resulting in stronger hyperfine coupling that supresses nuclear spin flip-flops, and thus dephasing. In contrast, the methoxy group in ZnL is farther from the ligand-centered electron, so nuclear spins exchange magnetization more freely, resulting in dephasing. We determine the spin pair that contributes most significantly to the dephasing dynamics in ZnL, which is a hydrogen pair on the methoxy group, and consider the induced dynamics from this single contribution for both molecules in Figure~\ref{fig:niZnMajorContributors}.
We can suppress the spin dephasing in the ZnL structure by considering two different substituted ligands, which replace the methoxy methyl group with CF$_3$ and ethylene C$_2$H$_3$, denoted L$_F$ and L$_E$ respectively, shown in Fig.~\ref{fig:schematic}~(b). These modifications change the hyperfine couplings: the L$_F$ modification entails a different gyromagnetic ratio and a slightly modified bond length, while the L$_E$ modification alters the connectivity of the nuclei. Table~\ref{tab:modifiedSpinPairData} shows the modulation depth $\alpha_{kl}^2$ corresponding to the possible interactions between the three hydrogen- or fluorine-spin pairs in all three ligands L, L$_E$, and L$_F$.
\begin{table}[h!]
    \centering
     \begin{tabular}{@{\extracolsep{4pt}}l c c c c c c} 
        \hline \hline 
        & \multicolumn{3}{c}{Ni} & \multicolumn{3}{c}{Zn} \\
        \cmidrule{2-4} \cmidrule{5-7}
          & L &  L$_E$ & L$_F$ &  L  & L$_E$ & L$_F$  \\ 
        \cline{2-2}\cline{3-3}\cline{4-4}\cline{5-5}\cline{6-6}\cline{7-7} 
1		& 2.7$\cdot 10^{-3}$ & 1.2$\cdot 10^{-3}$ & 5.0$\cdot$10$^{-6}$ & 0.11 & 8.8$\cdot 10^{-2}$ & 1.1$\cdot$10$^{-7}$ \rule{0pt}{2.2ex} \\
2		& 3.5$\cdot 10^{-5}$ & 5.6$\cdot$10$^{-5}$ & 1.5$\cdot 10^{-4}$ & 3.5$\cdot$10$^{-5}$ & 3.8$\cdot 10^{-3}$ & 1.4$\cdot 10^{-4}$ \\
3		& 4.1$\cdot 10^{-2}$ & 3.0$\cdot 10^{-2}$ & 6.8$\cdot 10^{-4}$ & 0.33 & 0.15 & 1.4$\cdot 10^{-3}$ \\
\hline\hline
    \end{tabular}
    \caption{Modulation depths for the three hydrogen or fluorine spin pairs. The indices 1, 2, and 3 represent the three possible interactions between the three spins on the original ligand L, the fluorine substituted ligand L$_F$, and the ethylene substituted ligand L$_E$.}
    \label{tab:modifiedSpinPairData}
\end{table}
The L$_F$ modification effectively removes the spins from contributing to electron dephasing, as the dipolar coupling between fluorines and hyperfine coupling to the electron is weaker, decreasing the modulation depth considerably. The L$_E$ modification also decreases the modulation depth for pairs 1 and 3, while increasing it slightly for pair 2. This discrepancy with pair 2 is likely due to the increased distance between the spin pair and the electron in both NiL and ZnL, moving spin pair 2 beyond the spin diffusion barrier. We compare the dynamics of these modified structures with the unaltered molecules for NiL and ZnL in Figures~\ref{fig:niModifiedSingleMol}~(a) and (b), respectively.
We note that in both the nickel and zinc structures the modifications improve the coherence in early times, likely due to the decrease in modulation depth for the spin pairs shown in Table~\ref{tab:modifiedSpinPairData}. The L$_F$ modifications improve early time coherence more than the L$_E$ modifications.

\begin{figure}[h!]
    \centering
    \includegraphics[width=0.8\linewidth]{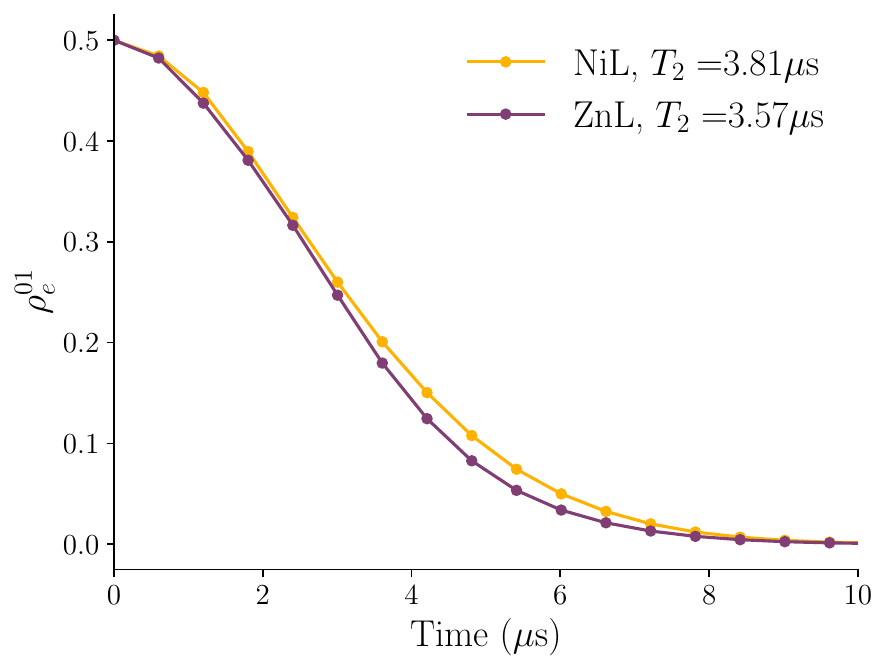}
    \caption{Electron spin coherence over time for NiL and ZnL in the presence of a hydrogen spin bath.}
    \label{fig:niZnFullHBath}
\end{figure}

In addition to the intramolecular spin effects, we now consider the role of the solvent spins in the electron dephasing dynamics. Details on the treatment of the solvent spin system are shown in S3 in the SI. Figure~\ref{fig:niZnFullHBath} shows the dynamics for NiL and ZnL with the intramolecular, molecule-solvent, and solvent-solvent spin pairs included, along with computed $T_2$ times. Hyperfine coupling between the electron spin and solvent nuclear spins is computed with the point-dipole approximation. Consistent with experimental results, NiL remains coherent longer than ZnL, and we observe the expected full decay of the coherence with decay constants of the same order of magnitude as experiment. 

Including all intramolecular and solvent interactions, we examine the \textcolor{black}{resulting $T_2$ coherence times} of the modified metal-ligand systems in \textcolor{black}{Table~\ref{tab:T2-times}}, which show\textcolor{black}{s} the \textcolor{black}{$T_2$ times} of all three variants of the zinc and nickel species, \textcolor{black}{ along with the experimentally measured times from Ref.~\citenum{Martins:2026}}.
\begin{table}[h!]
    \centering
     \begin{tabular}{@{\extracolsep{4pt}}l l c c c } 
        \hline \hline 
       \multicolumn{2}{c}{} ML$_E$ & ML$_F$& ML& ML Exp.~\cite{Martins:2026}\\
        \cmidrule{1-1} \cmidrule{2-3} \cmidrule{4-4} \cmidrule{5-5}
      Ni& 4.06 & 3.87 & 3.81 & 3.7\\
      Zn & 3.77& 3.69& 3.57& 1.78\\
    \hline\hline
    \end{tabular}
    \caption{\textcolor{black}{$T_2$ relaxation times ($\mu$s) for the original ligand L, the fluorine substituted ligand L$_F$, and the ethylene substituted ligand L$_E$, comparing theoretical results generated from the TCL2 method and experimental data from Ref.~\citenum{Martins:2026}.}}
    \label{tab:T2-times}
\end{table}

Similar to the isolated molecules, both modifications improve coherence in the ZnL and NiL structures. Again, the L$_F$ modification provides the greatest increase in coherence time for both complexes. 

\begin{figure}[h!]
    \centering
    \includegraphics[width=0.8\linewidth]{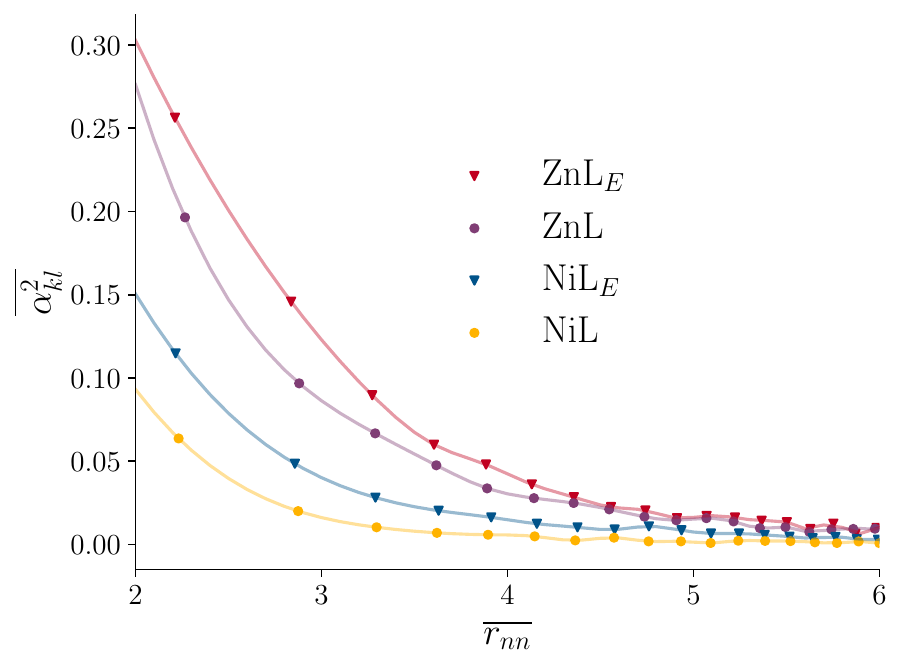}
    \caption{Modulation depth for hydrogen-spin pairs, where one spin is located on the CH$_3$ or C$_2$H$_3$ and the other in the solvent environment, versus the internuclear distance. The results have been averaged over all three nuclear spins present on the CH$_3$ or C$_2$H$_3$ group, and a spline fit is included for clarity.}
    \label{fig:niZnModMScomparison}
\end{figure}

To isolate the role of the molecule-solvent spin pairs on decoherence, we consider the average modulation depth due to the coupling of a hydrogen of the methoxy methyl or ethylene group with a solvent hydrogen, along with the average molecule-solvent internuclear distance $r_{nn}$ for NiL, ZnL, NiL$_E$ and ZnL$_E$ in Figure~\ref{fig:niZnModMScomparison}. We do not include the L$_F$ molecule-solvent interactions since heteronuclear spin pairs do not significantly contribute.~\cite{Krogmeier:2026,Liu:2007,Cywinski:2009,Chekhovich:2017}

\begin{figure}[h!]
    \centering
    \includegraphics[width=0.8\linewidth,  trim = 0cm 0cm 40cm 0cm, clip]{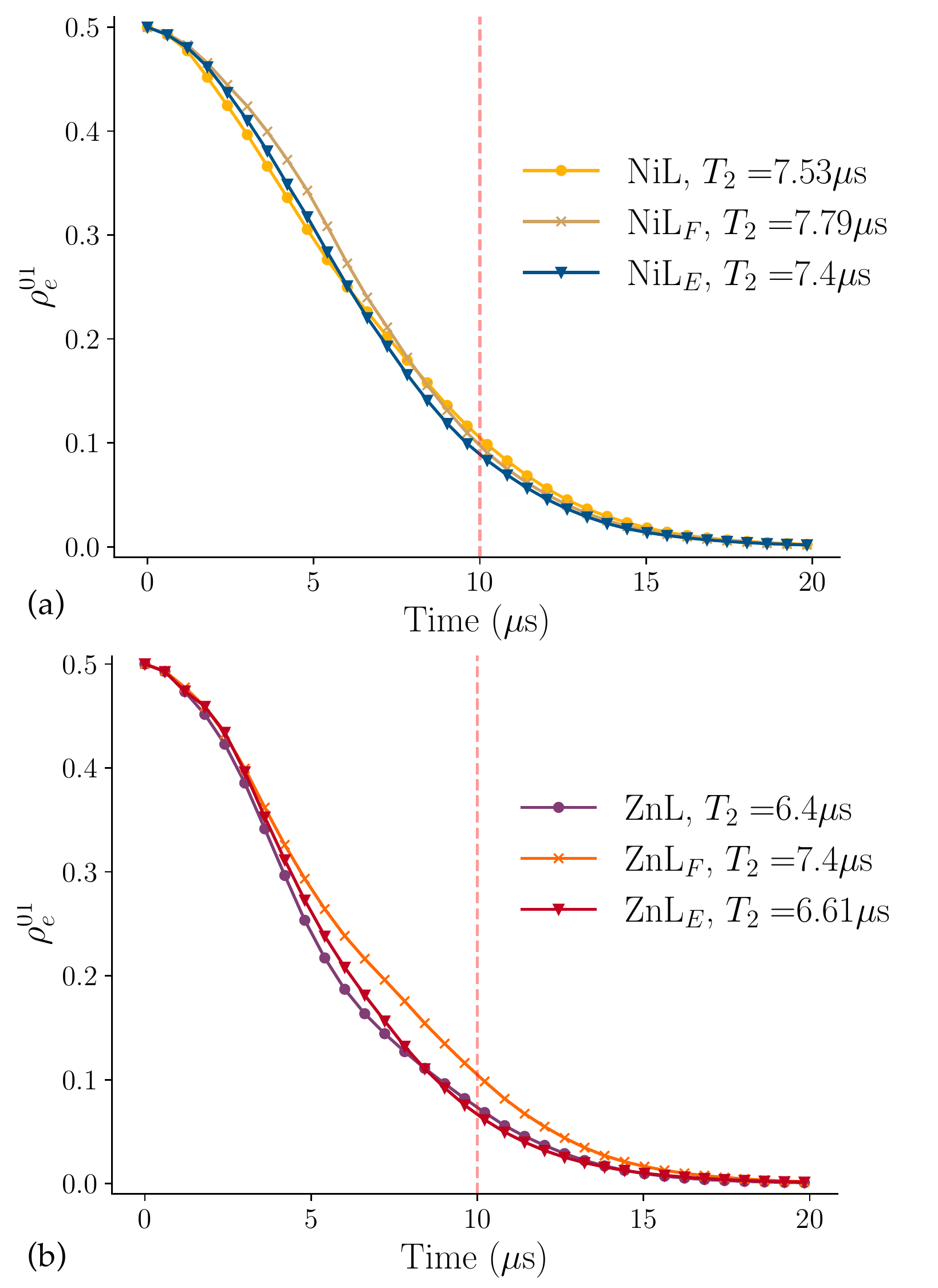}
    \caption{Electron coherence dynamics of (a) unmodified NiL along with modified complexes NiL$_F$ and NiL$_E$ and (b) unmodified ZnL along with modified complexes ZnL$_F$ and ZnL$_E$, all in the presence of a hydrogen bath reduced in density by 50\%. A vertical dotted red line is included at the time where the experimental data has completely dephased.}
    \label{fig:concentration-modification}
\end{figure}

Finally, to consider the effects of solvent-solvent pairs, we consider the dephasing dynamics in a solution with lower solvent concentration. Figure~\ref{fig:concentration-modification} shows the dynamics in a solvent with the hydrogen spin density reduced by 50\%, with the vertical dotted line indicating the complete dephasing timescale observed in experiment. Importantly, the $T_2$ time improves by a factor of two when removing spins from the solvent, e.g. by deuteration, which is known to increase coherence times; however, Fig.~\ref{fig:concentration-modification} demonstrates that at lower solvent-spin density, molecular modifications can also appreciably alter spin dephasing times.

\section{Discussion}
For the considered qubit candidates, our results show that molecular modifications can improve coherence via detuning intramolecular spin pairs or molecule-solvent pairs. The first modification, L$_F$, replaces the methoxy methyl group with a CF$_3$ group, removing three hydrogens entirely from the system, reducing both intramolecular and molecule-solvent spin pair contributions to dephasing,  compared to the original methyl group. This modification results in improved coherence times in both the isolated molecule and with the spin bath, shown in Figs.~\ref{fig:niModifiedSingleMol} and \textcolor{black}{Table ~\ref{tab:T2-times}}, respectively. 

The second modification, L$_E$, replaces the methoxy methyl group with a C$_2$H$_3$ group, and presents a more nuanced case. Figure~\ref{fig:niModifiedSingleMol} shows that the L$_E$ modification increases the coherence in both isolated molecules at early times, decreasing the contribution of the intramolecular spin pairs to electron-spin dephasing. On the other hand, Figure~\ref{fig:niZnModMScomparison} shows that for both metals the L$_E$ modification increases the modulation depth from molecule-solvent spin pairs, indicating an increased contribution from these spin pairs to electron-spin dephasing compared to the original methyl group. We hypothesize this is because the ethylene modification moves the molecular hydrogen spins further from the electron density in both systems, effectively moving the pair beyond the spin diffusion barrier and facilitating electron dephasing. \textcolor{black}{A greater discussion of the spin diffusion barrier is included in the SI.} Despite these competing effects, in both \textcolor{black}{Table.~\ref{tab:T2-times}}and Fig.~\ref{fig:concentration-modification} we see improvements in the spin coherence times with the L$_E$ modification compared to the unmodified structures. These improvements indicate that the decrease in the intramolecular spin-pair modulation depth is more significant than the increase in the molecule-solvent spin pair modulation depths.

\section{Conclusion}

By combining electronic structure calculations and the TCL2 master equation we qualitatively replicate experimental observations of variants of nickel and zinc molecular qubit candidates. We systematically delineate contributions to electron spin dephasing in terms of intramolecular, solvent-solvent, and molecule-solvent nuclear spin pairs. Our results confirm that dilution or elimination of solvent spins is the most effective way to increase the coherence time of molecular spins under these conditions. Nonetheless, we are able to understand experimentally observed differences in the coherence due to ligand substitutions. These substitutions result in subtle changes in the nuclear-spin effects, which depend on both the electronic (hyperfine) and nuclear (dipolar) couplings, and modulate the extent of the so-called spin-diffusion barrier. This complexity is challenging to understand without electronic structure information, because even small geometric changes can suppress or amplify decoherence due to nuclear spin pairs. \textcolor{black}{A key approximation at play is the point-dipole treatment of molecule-solvent and solvent-solvent spin pair contributions to the spin dephasing. Future research will be dedicated to testing the limits of this approximation, and developing improvements for more complex solvent structures and interactions.}

An important outcome highlighted by these results is that different molecular modifications can have competing impacts on electron-spin dephasing. This is seen explicitly with the L$_F$ and L$_E$ modifications, where the L$_F$ modification eliminates intramolecular spin pair contributions entirely in the timescale of interest and limits flip-flops with the solvent spins due to the difference in gyromagnetic ratios of solvent hydrogen spins and molecular fluorine spins. The L$_E$ modification decreases the intramolecular spin pair modulation depths, decreasing those contributions to spin dephasing; however this modification also enables molecule-solvent hydrogen spin pairs to continue dephasing the electron. We take these two modification schemes as evidence suggesting that molecular alterations to improve electron spin coherence times requires detailed analysis of all possible spin-spin interactions in order to accurately estimate the impact of a ligand substitution.

While generalizable design principles are challenging to extract, here we present a workflow for determining improvable components in low-temperature spin-spin induced dephasing dynamics. This workflow is suitable for the analysis of an electron dephasing due to interactions with a nuclear-spin environment, and could be extended more generally to molecular or material spins adsorbed onto surfaces or other sources of environmental nuclear spins. Our approach is scalable, \textcolor{black}{relies on a single electronic structure calculation,} and derived from microscopic principles, resulting in a practical theory that provides physical insight into complex molecular spin dynamics.

\section{Associated Content}

\subsection{Data Availability Statement}

The data and codes used in this study are freely available in the Zenodo online repository.~\cite{krogmeier:mqt}

\subsection{Supporting Information}
Details on the hydrogen spin bath, spin densities for the modified molecular structures, and the treatment of solvent-solvent spin pairs can be found in the Supporting Information.  

\section{Acknowledgements}

KHM acknowledges start-up funding from the University of Minnesota. This work was partially supported by the donors of ACS Petroleum Research Fund under Doctoral New Investigator Grant 68027-DNI6. K.H.-M. served as Principal Investigator on ACS PRF 68027-DNI6 that provided partial support for T.J.K. \textcolor{black}{and J.B.}

\section{References}

\bibliography{main} 

\end{document}